\documentclass[aps,prd,twocolumn,superscriptaddress,showpacs]{revtex4}
\usepackage[dvips]{graphicx}
\usepackage{amsmath,amssymb,times}

\newcommand{\bequ}{\begin{equation}}
\newcommand{\eequ}{\end{equation}}
\newcommand{\bea}{\begin{eqnarray}}
\newcommand{\eea}{\end{eqnarray}}

\DeclareSymbolFont{boldletters}{OML}{cmm} {b}{it}
\DeclareSymbolFontAlphabet{\mathbit}{boldletters}
\DeclareMathSymbol{\alpha}{\mathalpha}{letters}{"0B}
\DeclareMathSymbol{\beta}{\mathalpha}{letters}{"0C}
\DeclareMathSymbol{\gamma}{\mathalpha}{letters}{"0D}
\DeclareMathSymbol{\delta}{\mathalpha}{letters}{"0E}
\DeclareMathSymbol{\epsilon}{\mathalpha}{letters}{"0F}
\DeclareMathSymbol{\zeta}{\mathalpha}{letters}{"10}
\DeclareMathSymbol{\eta}{\mathalpha}{letters}{"11}
\DeclareMathSymbol{\theta}{\mathalpha}{letters}{"12}
\DeclareMathSymbol{\iota}{\mathalpha}{letters}{"13}
\DeclareMathSymbol{\kappa}{\mathalpha}{letters}{"14}
\DeclareMathSymbol{\lambda}{\mathalpha}{letters}{"15}
\DeclareMathSymbol{\mu}{\mathalpha}{letters}{"16}
\DeclareMathSymbol{\nu}{\mathalpha}{letters}{"17}
\DeclareMathSymbol{\xi}{\mathalpha}{letters}{"18}
\DeclareMathSymbol{\pi}{\mathalpha}{letters}{"19}
\DeclareMathSymbol{\rho}{\mathalpha}{letters}{"1A}
\DeclareMathSymbol{\sigma}{\mathalpha}{letters}{"1B}
\DeclareMathSymbol{\tau}{\mathalpha}{letters}{"1C}
\DeclareMathSymbol{\upsilon}{\mathalpha}{letters}{"1D}
\DeclareMathSymbol{\phi}{\mathalpha}{letters}{"1E}
\DeclareMathSymbol{\chi}{\mathalpha}{letters}{"1F}
\DeclareMathSymbol{\psi}{\mathalpha}{letters}{"20}
\DeclareMathSymbol{\omega}{\mathalpha}{letters}{"21}
\DeclareMathSymbol{\varepsilon}{\mathalpha}{letters}{"22}
\DeclareMathSymbol{\vartheta}{\mathalpha}{letters}{"23}
\DeclareMathSymbol{\varpi}{\mathalpha}{letters}{"24}
\DeclareMathSymbol{\varrho}{\mathalpha}{letters}{"25}
\DeclareMathSymbol{\varsigma}{\mathalpha}{letters}{"26}
\DeclareMathSymbol{\varphi}{\mathalpha}{letters}{"27}
\DeclareMathSymbol{\Gamma}{\mathalpha}{letters}{"00}
\DeclareMathSymbol{\Delta}{\mathalpha}{letters}{"01}
\DeclareMathSymbol{\Theta}{\mathalpha}{letters}{"02}
\DeclareMathSymbol{\Lambda}{\mathalpha}{letters}{"03}
\DeclareMathSymbol{\Xi}{\mathalpha}{letters}{"04}
\DeclareMathSymbol{\Pi}{\mathalpha}{letters}{"05}
\DeclareMathSymbol{\Sigma}{\mathalpha}{letters}{"06}
\DeclareMathSymbol{\Upsilon}{\mathalpha}{letters}{"07}
\DeclareMathSymbol{\Phi}{\mathalpha}{letters}{"08}
\DeclareMathSymbol{\Psi}{\mathalpha}{letters}{"09}
\DeclareMathSymbol{\Omega}{\mathalpha}{letters}{"0A}

\begin{document}
\title{
Quark-mass dependence of three-flavor QCD phase diagram at zero and imaginary chemical potential: Model prediction
}

\author{Takahiro Sasaki}
\email[]{sasaki@phys.kyushu-u.ac.jp}
\affiliation{Department of Physics, Graduate School of Sciences, Kyushu University,
             Fukuoka 812-8581, Japan}

\author{Yuji Sakai}
\email[]{sakai@phys.kyushu-u.ac.jp}
\affiliation{Department of Physics, Graduate School of Sciences, Kyushu University,
             Fukuoka 812-8581, Japan}

\author{Hiroaki Kouno}
\email[]{kounoh@cc.saga-u.ac.jp}
\affiliation{Department of Physics, Saga University,
             Saga 840-8502, Japan}

\author{Masanobu Yahiro}
\email[]{yahiro@phys.kyushu-u.ac.jp}
\affiliation{Department of Physics, Graduate School of Sciences, Kyushu University,
             Fukuoka 812-8581, Japan}

\date{\today}

\begin{abstract}
We draw the three-flavor phase diagram 
as a function of light- and strange-quark masses 
for both zero and imaginary quark-number chemical potential, using 
the Polyakov-loop extended Nambu--Jona-Lasinio model with 
an effective four-quark vertex depending on the Polyakov loop. 
The model prediction is qualitatively consistent with 2+1 flavor lattice QCD prediction 
at zero chemical potential and with degenerate three-flavor lattice QCD 
prediction at imaginary chemical potential. 
\end{abstract}

\maketitle

\begin{figure}[htbp]
\begin{center}
\hspace{-10pt}
 \includegraphics[width=0.24\textwidth]{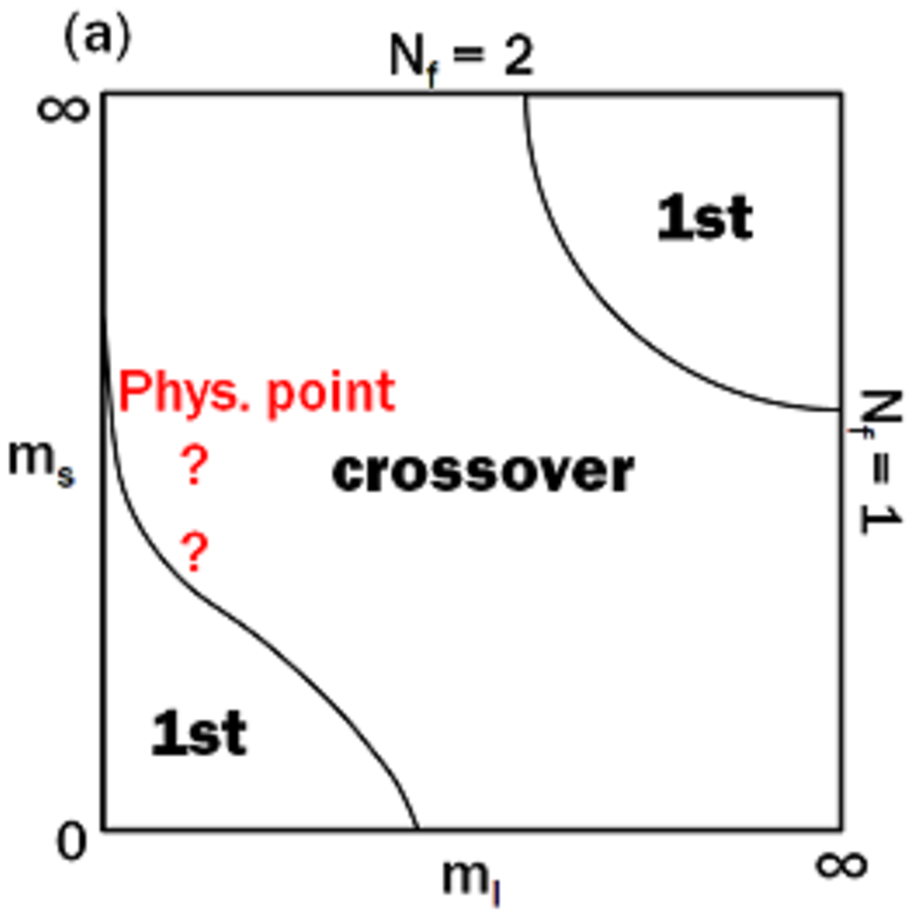} 
 \includegraphics[width=0.24\textwidth]{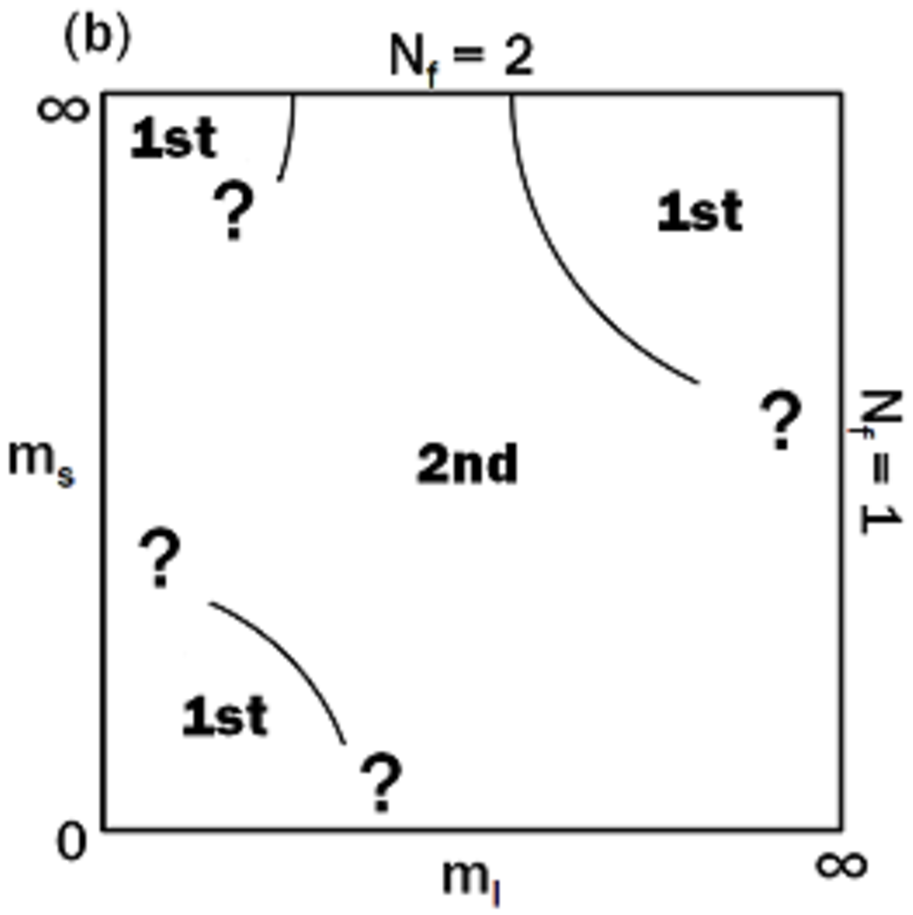} 
\vspace{-15pt}
\end{center}
\caption{Sketch of the three-flavor phase diagram 
in the $m_{l}$-$m_{s}$ plane. 
Panel (a) shows a sketch for the chiral transition at $\mu=0$. 
The solid line denotes the second-order chiral transition line. 
Panel (b) shows a sketch for the RW transition at the end point 
$(T,\theta)=(T_{\rm RW},\pi)$. 
The solid line means the boundary between the first- and 
second-order transition regions. Panel (b) is based on 
two-flavor~\cite{D'Elia-3} and degenerate three-flavor~\cite{FP2010} 
LQCD results 
that the RW transition at the end point is first order for light and 
heavy quark masses, but second order for intermediate masses. 
}
\label{3f-diagram}
\end{figure}

{\it Introduction.} 
Determination of the order of QCD phase transitions is an important 
subject not only in hadron physics but also in cosmology~\cite{YAoki}. 
The chiral and deconfinement transitions are 
widely believed to be 
crossover at zero chemical potential, when physical 
values are taken for light and strange quark masses, 
$m_{l}$ and $m_{s}$~\cite{Borsanyi,Soeldner,Kanaya}. 
However, the order of the transitions is sensitive 
to the number ($N_{f}$) of flavors and the values of $m_{l}$ and $m_{s}$. 
A sketch of the three-flavor phase diagram is plotted in 
Fig.~\ref{3f-diagram}(a) as a function of $m_{l}$ and $m_{s}$ 
for the case of zero chemical potential ($\mu$). 
This sketch, sometimes called the Columbia plot, is based on theoretical 
considerations and lattice QCD (LQCD) data~\cite{Laermann,Kanaya,FP2007}. 
The physical point lies near the second-order transition (solid) line.

For higher $\mu$, the chiral crossover at the physical point is expected to 
become first order. In this case, 
there appears a critical end point (CEP) 
of the first-order transition line, and 
the transition becomes second order on CEP~\cite{AY,Barducci_CEP,Kashiwa}. 
However, clear evidence of the behavior is not shown yet 
by LQCD because of the sign problem at real $\mu$.

On the contrary, LQCD is feasible~\cite{FP,D'Elia,Chen34,Chen,D'Elia-iso,Cea,D'Elia-3,FP2010,Nagata,Takaishi} at imaginary $\mu=i\theta T$, 
where $T$ stands for the temperature and $\theta$ represents 
the dimensionless chemical potential. 
QCD has a periodicity of $2\pi/3$ in $\theta$ called 
the Roberge-Weiss (RW) periodicity~\cite{RW}, because QCD is invariant
under the extended ${\mathbb Z}_3$ transformation~\cite{Sakai}. 
At $\theta =\pi/3$ mod $2\pi/3$, 
there appears a first-order transition at $T$ higher than some 
temperature $T_{\rm RW}$~\cite{RW}. 
This is now called the RW transition~\cite{RW}. 
On the RW transition line starting from 
the end point $(\theta,T)=(\pi ,T_{\rm RW})$, 
a spontaneous $C$ symmetry breaking ~occurs~\cite{Kouno}. 
Very recently, the order of the $C$ symmetry breaking at the RW end point has 
been analyzed by two-flavor~\cite{D'Elia-3} and 
degenerate three-flavor~\cite{FP2010} LQCD. 
For the two cases, the order is first order at small and large 
quark masses, but second order for intermediate masses. 
Figure \ref{3f-diagram}(b) is a sketch based on the LQCD results 
for the RW phase transition at the end point . 
Most of the region is unknown at the present stage.

As an approach complementary to first-principle LQCD, 
we can consider effective models such as the  Nambu--Jona-Lasinio (NJL) 
model~\cite{AY,Kashiwa,Reinberg} and 
the Polyakov-loop extended Nambu--Jona-Lasinio (PNJL) 
model~\cite{Kouno,Sakai,Meisinger,Fukushima,Ratti,Rossner,Schaefer,Kashiwa1,Sakai2,Kashiwa5,Matsumoto,Sasaki-T,Sakai5,Gatto}. 
The NJL model can describe the chiral symmetry breaking, 
but not the confinement mechanism. 
The PNJL model is designed to make it possible 
to treat both the mechanisms. 
The effective models have ambiguity in determining 
their parameters~\cite{Kashiwa5,Sakai2,Matsumoto}.  
We then take the following strategy. 
We first construct an effective model and determine 
parameters of the model in the regions where LQCD is feasible. 
Next, we predict physical quantities 
in the regions where LQCD is not feasible, using the constructed model.

The original PNJL model cannot reproduce LQCD data 
at imaginary $\mu$ quantitatively~\cite{Sakai2}. 
This shortcoming of the PNJL model seems to be originated 
in the fact that the correlation between the chiral condensate $\sigma$
and the Polyakov loop $\Phi$ is too weak.   
Therefore, in Ref.~\cite{Sakai5}, we extended the two-flavor PNJL model 
by introducing the effective four-quark vertex depending on 
$\Phi$. This effective vertex includes additional mixing effects 
between $\sigma$ and $\Phi$. 
The new model is called the entanglement PNJL (EPNJL) model. 
The two-flavor EPNJL model reproduces LQCD data 
at zero and imaginary $\mu$, particularly 
on strong correlations between the chiral and deconfinement transitions 
and also on quark-mass dependence of the order 
of the RW end point~\cite{D'Elia-3}. 
The two-flavor EPNJL model reproduces all LQCD data, 
without changing the parameters, at small real $\mu$ 
without \cite{Sakai5} and with strong magnetic field~\cite{Gatto}
and at finite isospin chemical potential~\cite{Sakai5}.

In this paper, we extend the two-flavor EPNJL model to the three-flavor case. 
Parameters of the three-flavor EPNJL model are determined from 
LQCD data at zero $\mu$ and at the RW end point. 
The Columbia pot is drawn for the chiral transition at zero $\mu$ and 
for the $C$ symmetry breaking at the RW end point.

{\it Model setting.} 
We start with the three-flavor PNJL model. The Lagrangian density of 
the model is  
\begin{align}
 {\cal L}  
=& {\bar q}(i \gamma_\nu D^\nu - {\hat m_0} )q  
  + G_{\rm S} \sum_{a=0}^{8} 
    [({\bar q} \lambda_a q )^2 +({\bar q }i\gamma_5 \lambda_a q )^2] 
\nonumber\\
 &- G_{\rm D} \Bigl[\det_{ij} {\bar q}_i (1+\gamma_5) q_j 
           +\det_{ij} {\bar q}_i (1-\gamma_5) q_j \Bigr]
\nonumber\\
&-{\cal U}(\Phi [A],{\bar \Phi} [A],T) , 
\label{L}
\end{align} 
where $D^\nu=\partial^\nu + iA^\nu=\partial^\nu +i\delta^{\nu}_{0}gA^0_a{\lambda_a / 2}$ with the gauge coupling $g$ and the Gell-Mann matrices $\lambda_a$. 
Three-flavor quark fields $q=(q_u,q_d,q_s)$ have current quark masses 
${\hat m_0}={\rm diag}(m_u,m_d,m_s)$. 
In the interaction part, $G_{\rm S}$ and $G_{\rm D}$ denote coupling constants 
of the scalar-type four-quark and the Kobayashi-Maskawa-'t Hooft (KMT) determinant interaction~\cite{KMK,tHooft}, 
respectively, in which the determinant 
runs in the flavor space. 
The KMT determinant interaction breaks 
the $U_\mathrm{A} (1)$ symmetry explicitly.

In the PNJL model, the gauge field $A_\mu$ is treated as a homogeneous and static background field~\cite{Fukushima}. 
The Polyakov loop $\Phi$ and its conjugate $\Phi ^*$ are determined in the Euclidean space by
\begin{align}
\Phi &= {1\over{3}}{\rm tr}_{\rm c}(L),
~~~~~\Phi^* ={1\over{3}}{\rm tr}_{\rm c}({\bar L}),
\label{Polyakov}
\end{align}
where $L  = \exp(i A_4/T)$ with $A_4=iA_0$. 
In the Polyakov gauge, $A_4$ is diagonal in the color space. 
The Polyakov potential $\mathcal{U}$ is assumed to be a function of $\Phi$ and $\Phi^*$.
We take the Polyakov potential of Ref.~\cite{Rossner}: 
\begin{align}
&{\cal U} = T^4 \Bigl[-\frac{a(T)}{2} {\Phi}^*\Phi\notag\\
      &~~~~~+ b(T)\ln(1 - 6{\Phi\Phi^*}  + 4(\Phi^3+{\Phi^*}^3)
            - 3(\Phi\Phi^*)^2 )\Bigr] ,
            \label{eq:E13}\\
&a(T)   = a_0 + a_1\Bigl(\frac{T_0}{T}\Bigr)
                 + a_2\Bigl(\frac{T_0}{T}\Bigr)^2,~~~~
b(T)=b_3\Bigl(\frac{T_0}{T}\Bigr)^3 .
            \label{eq:E14}
\end{align}
Parameters of $\mathcal{U}$ are determined to reproduce LQCD data at finite $T$ in the pure gauge limit.

Using the mean field approximation to the quark-quark interactions 
in \eqref{L}, one can get the thermodynamic potential 
(per volume)~\cite{Matsumoto}: 
\begin{align}
\Omega
&= -2 \sum_{i=u,d,s} \int \frac{d^3 \vec{p}}{(2\pi)^3}
   \Bigl[ N_\mathrm{c} E_{\vec{p},f} \nonumber\\
&        + \frac{1}{\beta}
           \ln~ [1 + 3(\Phi+\Phi^* e^{-\beta (E_{\vec{p},i}-\mu)}) e^{-\beta (E_{\vec{p},i}-\mu)} \notag\\
&           + e^{-3\beta (E_{\vec{p},i}-\mu)}] \notag\\
&        + \frac{1}{\beta} 
           \ln~ [1 + 3(\Phi^*+\Phi e^{-\beta (E_{\vec{p},i}+\mu)}) e^{-\beta (E_{\vec{p},i}+\mu)} \notag\\
&           + e^{-3\beta (E_{\vec{p},i}+\mu)}]
	      \Bigl]
	      \nonumber\\
&+ \Bigl( \sum_{i=u,d,s}  2 G_{\rm S} \sigma_{ii}^2 
 - 4 G_{\rm D} \sigma_{uu}\sigma_{dd}\sigma_{ss} \Bigr) \notag\\
& +{\cal U}(\Phi [A],\Phi^* [A],T), 
\label{PNJL-Omega}
\end{align}
where $\sigma_{ij} \equiv \langle {\bar q}_i q_j \rangle$
and $E_{\vec{p}}^i \equiv \sqrt{\vec{p}{}^2+{M_{ii}}^2}$ for $i,j=u,d,s$. 
The dynamical quark mass $M_{ii}$ is defined by 
\bea
M_{ii}=m_{i}-4G_{\rm S}\sigma_{ii}+  2G_{\rm D} \sigma_{jj} \sigma_{kk} 
\eea
for $i \neq j \neq k$. 
The variables $\Phi$, ${\Phi}^*$, 
$\sigma_l(\equiv \sigma_{uu} = \sigma_{dd})$ and $\sigma_{s}(\equiv\sigma_{ss})$ 
are determined by the stationary condition~\cite{Matsumoto}.

When $\mu=i\theta T$, the thermodynamic potential of QCD 
has the RW periodicity~\cite{RW}, i.e. a periodicity of $2\pi/3$ in $\theta$. 
The PNJL thermodynamic potential $\Omega$ 
of \eqref{PNJL-Omega} also has this periodicity, 
since the potential is invariant 
under the extended ${\mathbb Z}_3$ transformation~\cite{Sakai}. 
At $\theta =0$ and $\pi$, $\Omega$ is $C$ symmetric~\cite{Kouno}. 
Particularly at $\theta =\pi $, it is spontaneously 
broken at higher $T$~\cite{RW,Kouno}. 
The order parameter of the spontaneous $C$ symmetry breaking
is a $\theta$-odd quantity such as 
the imaginary part of the modified Polyakov loop $\Psi=\Phi e^{i \theta}$~\cite{Kouno}.

An origin of the four-quark vertex $G_{\rm S}$ is a gluon exchange 
between quarks and its higher-order diagrams. 
If the gluon field $A_{\nu}$ has a vacuum expectation value 
$\langle A_{0} \rangle$ in its time component, 
$A_{\nu}$ is coupled to $\langle A_{0} \rangle$ that is 
related to $\Phi$ through $L$~\cite{Kondo}.
Hence, $G_{\rm S}$ is changed into an effective vertex 
$G_{\rm S}(\Phi)$ depending on $\Phi$~\cite{Kondo}. 
Here, the effective vertex $G_{\rm S}(\Phi)$ 
is called the entanglement vertex and 
all interactions including $G_{\rm S}(\Phi)$ are referred to as 
the entanglement interactions. 
It is expected that $\Phi$ dependence of $G_{\rm S}(\Phi )$ 
will be determined 
in the future by the accurate method such as the exact renormalization group 
method~\cite{Braun,Kondo,Wetterich}. 
In this paper, however, we simply assume the following $G_{\rm S}(\Phi )$ that 
preserves the chiral symmetry, the $C$ symmetry~\cite{Kouno} 
and the extended $\mathbb{Z}_3$ symmetry~\cite{Sakai}: 
\begin{eqnarray}
G_{\rm S}(\Phi)=G_{\rm S}[1-\alpha_1\Phi\Phi^*-\alpha_2(\Phi^3+\Phi^{*3})]. 
\label{entanglement-vertex}
\end{eqnarray}
This modification changes the mesonic terms having $G_{\rm S}\sigma_{ii}$ 
and the dynamical quark masses $M_{ii}$ in $\Omega$.  
This is the three-flavor version of the EPNJL model, and 
this model has entanglement interactions in $G_{\rm S}\sigma_{ii}$ and 
$M_{ii}$ in addition to the covariant derivative included 
in the original PNJL model.
In principle, $G_{\rm D}$ can depend on $\Phi$, too. 
However, we found that the $\Phi$ dependence of $G_{\rm D}$ 
yields qualitatively the same effect on the phase diagram as that of 
$G_{\rm S}$. As a simple setup, we then neglect 
the $\Phi$ dependence of $G_{\rm D}$. 
In the present analysis, thus, the $\Phi$-dependence of $G_{\rm D}$ 
is renormalized in that of $G_{\rm S}$.

In the thermodynamic potential \eqref{PNJL-Omega},
we impose the isospin symmetry for the ${u}$-${d}$ sector 
($m_{l} \equiv m_{u}=m_{d}$) and take the three-dimensional 
cutoff $\Lambda$ for the momentum integration~\cite{Matsumoto}, 
because this model is nonrenormalizable. 
Hence, the three-flavor PNJL model has five parameters 
$G_{\rm S}$, $G_{\rm D}$, $m_l$, $m_s$, and $\Lambda$.
We use the parameter set of Table \ref{Table_NJL}~\cite{Reinberg}. 
These parameters are fitted to reproduce 
empirical values of $\pi$-meson mass and decay constant, 
$K$-meson mass and decay constant and $\eta'$ meson mass at vacuum. 

\begin{table}[h]
\begin{center}
\begin{tabular}{llllll}
\hline
~~$m_l(\rm MeV)$~~&~~$m_s(\rm MeV)$~~&~~$\Lambda(\rm MeV)$~~~&~~$G_{\rm S} \Lambda^2$
~~&~~$G_{\rm D}(0) \Lambda^5$~~
\\
\hline
~~~~~~~~5.5 &~~~~~~~~140.7 &~~~~~~~~602.3 &~~~~~~1.835 &~~~~~~12.36 &~~~~\\
\hline
\end{tabular}
\caption{
Summary of the parameter set in the NJL sector~\cite{Reinberg}. 
\label{Table_NJL}
}
\end{center}
\end{table}

Parameters of ${\cal U}$ are determined to reproduce 
LQCD data at finite $T$ in the pure gauge limit~\cite{Rossner}. 
The original value of $T_0$ is 270~MeV, but 
the deconfinement temperature $T_c$ determined by the EPNJL model 
with this value of $T_0$ is much larger than $T_c\approx 160$~MeV 
predicted by full LQCD \cite{Borsanyi,Soeldner,Kanaya}. 
Therefore, we rescale $T_0$ to 150~MeV so that the EPNJL model can reproduce 
$T_c=160$~MeV.

The parameters $\alpha_1$ and $\alpha_2$ in \eqref{entanglement-vertex} 
are so determined as to reproduce two results of LQCD at finite $T$. 
The first is a result of 2+1 flavor LQCD at $\mu=0$~\cite{YAoki} 
that the chiral transition 
is crossover at the physical point. The second is 
a result of degenerate three-flavor LQCD at $\theta=\pi$~\cite{FP2010} 
that the order of the RW end point is 
first order for small and large quark masses but second order 
for intermediate quark masses. 
The parameter set $(\alpha_1, \alpha_2)$ satisfying these conditions 
is located in the triangle region 
\bea
\{-1.5\alpha_1+0.3 < \alpha_2 <-0.86\alpha_1+0.32,~\alpha_2 >0\}. 
\label{triangle}
\eea
Here, we take $\alpha_1=0.25,~\alpha_2=0.1$ as a typical example.

\begin{figure}[htbp]
\begin{center}
\hspace{-10pt}
 \includegraphics[width=0.24\textwidth,bb=80 50 205 176,clip]{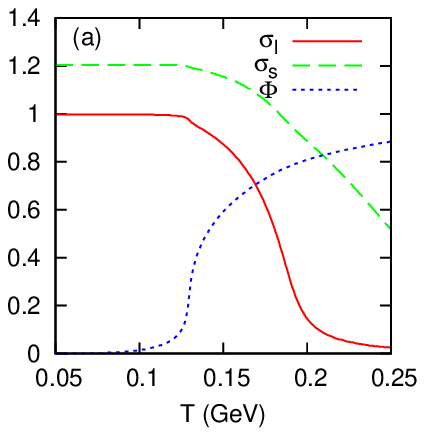} 
 \includegraphics[width=0.24\textwidth,bb=80 50 205 176,clip]{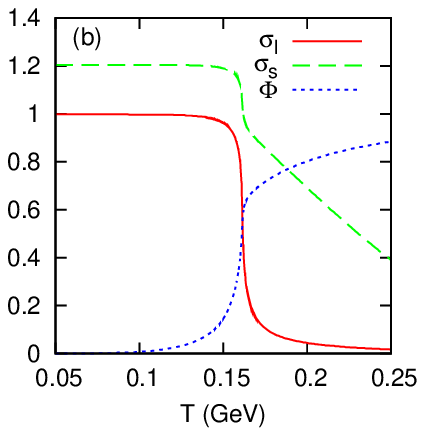}
\end{center}
\vspace{-10pt}
\caption{
$T$ dependence of the light- and strange-quark condensates 
and the Polyakov loop at $\mu =0$. 
The quark condensates are normalized by $\sigma_l=-0.0142~[{\rm GeV}^3]$ 
at $T=\mu =0$.  
Panels (a) and (b) represent results of the PNJL and EPNJL models, 
respectively. 
}
\label{order-param}
\end{figure}

{\it Results.} 
Figure~\ref{order-param} shows $T$ dependence of 
light- and strange-quark condensates, $\sigma_{l}$ and $\sigma_{s}$, and 
the Polyakov loop $\Phi$ at $\mu =0$. 
In the PNJL model of panel (a), 
$\sigma_{l}$ and $\sigma_{s}$ rapidly decrease at $T \approx 180$~MeV 
as $T$ increases, after $\Phi$ rapidly increases at $T \approx 130$~MeV 
as $T$ increases. 
Thus, the pseudocritical temperature 
of the chiral crossover is much higher than that of the deconfinement 
crossover. The same property is also seen 
in the two-flavor case~\cite{Sakai2}.    
In the EPNJL model of panel (b), meanwhile, 
the pseudocritical temperatures 
of the chiral and the deconfinement crossover almost coincide 
at $T \approx 160$~MeV.

\begin{figure}[htbp]
\begin{center}
\hspace{-10pt}
 \includegraphics[width=0.4\textwidth,bb=65 55 205 176,clip]{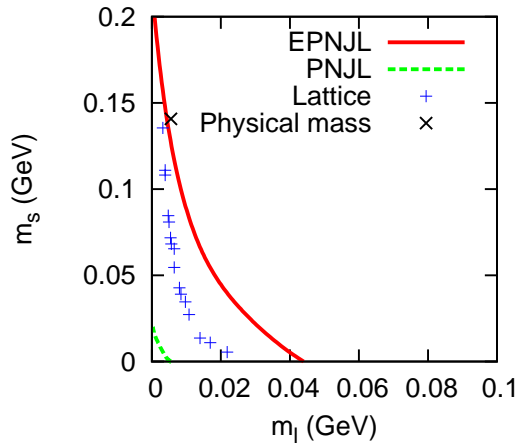}
\vspace{-15pt}
\end{center}
\caption{The order of the chiral transition 
in the $m_{l}$-$m_{s}$ plane at $\mu=0$. 
Solid and dotted lines and + symbols represent 
the second-order chiral-transition lines predicted by 
the PNJL and EPNJL models and LQCD~\cite{FP2007}), respectively. 
}
\label{col-mu0}
\end{figure}

Figure~\ref{col-mu0} shows the order of the chiral transition 
in the $m_{l}$-$m_{s}$ plane at $\mu=0$. 
This figure corresponds to the small $m_{l}$ and $m_{s}$ part of 
Fig.\ref{3f-diagram}(a). 
The second-order chiral-transition line is drawn 
for three cases, the PNJL result (dotted line) and 
the EPNJL result (solid line) and LQCD data (+ symbols)~\cite{FP2007}. 
For each of the three cases, there are the first-order region below the 
second-order line and the crossover region above the line. 
The second-order line predicted by the EPNJL model is close to that by 
LQCD data particularly near the physical point. Meanwhile, 
the first-order region predicted by the PNJL model is much smaller than 
that by LQCD data. 
Thus, the EPNJL model yields much better agreement with LQCD prediction than 
the PNJL model.

The deconfinement transitions predicted by the PNJL and EPNJL models are 
crossover in the whole region shown 
in Fig.~\ref{col-mu0}. 
In the EPNJL model, the crossover deconfinement transition 
almost coincides with the chiral transition, 
even if the chiral transition is crossover.

\begin{figure}[htbp]
\begin{center}
 \includegraphics[width=0.4\textwidth,bb=45 50 215 165,clip]{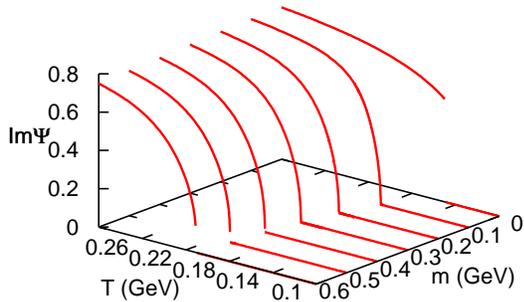} 
\end{center}
\vspace{-30pt}
\caption{ The imaginary part 
of the modified Polyakov loop at $\theta=\pi$ in the 
$m_{l}$-$T$ plane predicted by the EPNJL model with $m=m_l=m_s$.
}
\label{m-dep-RW}
\end{figure}

Now we consider the $C$ symmetry breaking at $\theta=\pi$ for 
the case of three degenerate flavors ($m_s=m_l$). 
Figure~\ref{m-dep-RW} represents the imaginary part of $\Psi$ 
as a function of $m_l$ and $T$  predicted by the three-flavor EPNJL model. 
When $m_l$ is large, the system is close to the pure gauge limit 
and hence the $C$-symmetry breaking is first order.  
When $m_l$ is small, meanwhile, the system is nearly 
chirally symmetric and therefore the transition is first order. 
In the intermediate mass region, the transition is second order. 
The result is consistent with the LQCD data~\cite{FP2010}.   

\begin{figure}[htbp]
\begin{center}
\hspace{-10pt}
 \includegraphics[width=0.4\textwidth,bb=73 55 205 190,clip]{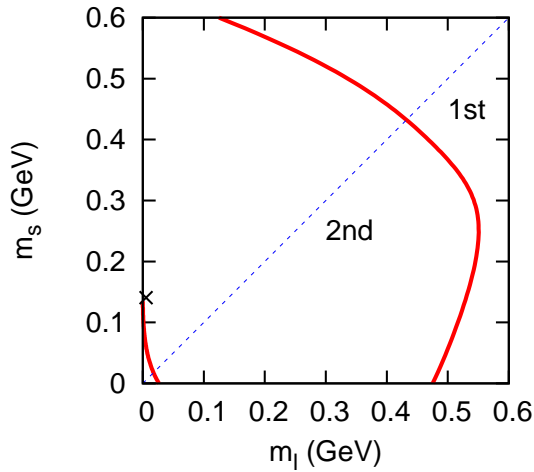}
\end{center}
\vspace{-20pt}
\caption{ The order of $C$ symmetry breaking 
at the RW end point predicted by the EPNJL model. 
The transition is first order below (above) the lower (upper) line, while 
it is second order between the two lines. 
The dotted line stands for a line of $m_l=m_s$, that is, 
the case of three degenerate flavors, 
whereas the $\times$ symbol means the physical mass.
}
\label{col-RW}
\end{figure}

Figure~\ref{col-RW} shows the phase diagram for the $C$-symmetry breaking 
at the RW end point predicted by the EPNJL model. 
The diagram is plotted as a function of $m_l$ and $m_s$ up to 
$\Lambda$, the upper limit for the present model to be applicable. 
The two solid lines represent boundaries between the first- and second-order 
transition regions. 
Below (above) the lower (upper) boundary, the transition is first order. 
The dotted line of $m_l=m_s$ corresponds to the case of $N_{f}=3$. 
On the dotted line, the order is first order for small and large 
masses but second order for intermediate masses, as expected. 
At the physical point, the order is second order for 
the present parameter set. However, 
the order can becomes first order at the physical point, 
if we take other parameter sets belonging to the region \eqref{triangle}. 
In the PNJL model, meanwhile, the transition is 
always first order in the entire region of the $m_l$-$m_s$ plane.

In Figs.~\ref{m-dep-RW} and ~\ref{col-RW}, the EPNJL prediction is 
shown for small and large current quark masses 
$m_q$ ($q=l,s$). The applicability of the NJL-type model 
to large $m_q$, however, is an open question. 
In fact it was pointed out that $m_q$ dependence of the chiral-transition temperature is not consistent with the corresponding LQCD 
results~\cite{Dumitru,Braun2}; 
as $m_q$ increases, the chiral-transition temperature goes up sizably 
in the NJL-type model but hardly changes in the LQCD results.   
In the EPNJL model, the chiral-transition temperature almost coincides with the deconfinement one that hardly depends on $m_q$, so that 
the EPNJL result is consistent with the LQCD result 
for the transition temperature.  
It was also pointed out that for large $m_q$ the pion mass $m_\pi$ 
calculated with the NJL-type model is larger than the corresponding 
LQCD result~\cite{Kahara}. 
In the NJL-type model the hadron mass calculation is questionable 
for large $m_q$, particularly when 
the calculated hadron mass is bigger than the cutoff $\Lambda$.   
Therefore, the ENJL predictions shown in Fig.~\ref{m-dep-RW} and \ref{col-RW} 
should be regarded as qualitative ones for the $m_q>100$MeV region 
where the calculated pion mass is bigger than $\Lambda$. 
However, the fact that there is the second-order region 
at intermediate $m_q$ ($<100$MeV) shows that there exists a boundary 
between the first- and second-order regions at large $m_q$. 
In this qualitative sense, the phase diagram of Fig.~\ref{col-RW} 
is reasonable for large $m_q$.

\begin{figure}[htbp]
\begin{center}
\hspace{-10pt}
 \includegraphics[width=0.24\textwidth,bb=67 50 207 176,clip]{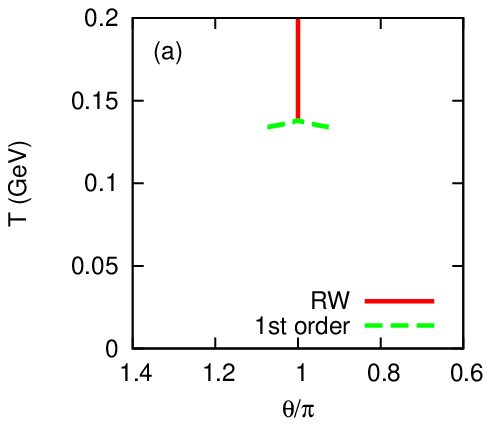}
 \includegraphics[width=0.24\textwidth,bb=67 50 207 176,clip]{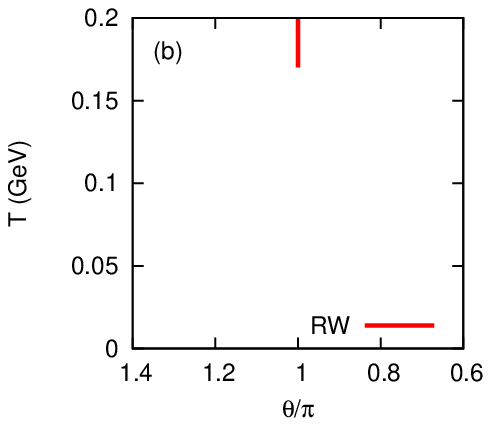} 
\hspace{-10pt}
\end{center}
\caption{
The phase diagram in the $\theta$-$T$ plane 
predicted by (a) the PNJL model and (b) the EPNJL model.
Here, physical values of $m_l$ and $m_s$ are 
taken. The solid line stands for the first-order RW transition line, while 
the dashed line corresponds to the first-order deconfinement line. 
}
\label{phase}
\end{figure}

Figure~\ref{phase} presents the phase diagram in the $\theta$-$T$ plane 
predicted by the PNJL and  EPNJL models, where $m_l$ and $m_s$ have 
physical values. 
In the PNJL model of panel (a),
a first-order RW transition (solid) line is connected at 
the RW end point to two first-order deconfinement (dashed) lines. 
Hence, the RW end point is a triple point. 
In the EPNJL model of panel (b), 
the RW transition is second order at the end point, so that 
there is no first-order deconfinement line 
connected to the first-order RW transition line. 
For other parameter sets in the parameter region \eqref{triangle}, 
the transition is weak first-order at the end point and 
hence the first-order RW transition line is connected at the RW 
end point to two very-short first-order deconfinement lines. 

{\it Summary.} 
In summary, we have extended the three-flavor PNJL model by introducing 
an entanglement vertex $G_{\rm S}(\Phi)$. 
The entanglement PNJL (EPNJL) model is consistent with 
2+1 flavor LQCD data for the chiral transition at $\mu=0$ and 
degenerate three-flavor LQCD data for the RW transition at the end point 
calculated very lately. 
The three-flavor phase diagram for the RW transition at the end point 
is first drawn in 
the $m_l$-$m_s$ plane by the EPNJL model justified above.

\noindent
\begin{acknowledgments}
The authors thank A. Nakamura, T. Saito, K. Nagata, and K. Kashiwa for useful discussions. 
H.K. also thanks M. Imachi, H. Yoneyama, H. Aoki, and M. Tachibana for useful discussions. 
T.S and Y.S. are supported by JSPS.
\end{acknowledgments}


\end{document}